\documentstyle[12pt]{article}
\begin{document}

\begin{center}

{\Large \bf Statistical Analysis of Magnetic Field Spectra}

\bigskip

Jian Wang

\bigskip

{\it
Department of Physics, \\
The University of Hong Kong,\\
Pokfulam Road, Hong Kong.
}

\bigskip

Hong Guo

\bigskip

{\it
Centre for the Physics of Materials,\\
Department of Physics, McGill University,\\
Montreal, Quebec, Canada H3A 2T8.
}

\end{center}

\vfill

\baselineskip 15pt               

We have calculated and statistically analyzed the magnetic-field 
spectrum (the ``B-spectrum'') at fixed electron Fermi energy for two
quantum dot systems with classically chaotic shape. This is a new
problem which arises naturally in transport measurements where
the incoming electron has a fixed energy while one tunes the magnetic
field to obtain resonance conductance patterns. The ``B-spectrum'', 
defined as the collection of values $\{B_i\}$ at which conductance 
$g(B_i)$ takes extremal values, is determined by a quadratic eigenvalue 
equation, in distinct difference to the usual linear eigenvalue problem
satisfied by the energy levels. We found that the lower part of the 
``B-spectrum'' satisfies the distribution belonging to Gaussian Unitary 
Ensemble, while the higher part obeys a Poisson-like behavior. 
We also found that the ``B-spectrum'' fluctuations of the chaotic 
system are consistent with the results we obtained from random matrices.

\vfill

\baselineskip 16pt

{PACS number: 72.10.Bg, 73.20.Dx, 73.40.Gk, 73.40.Lq}

\newpage

Due to recent advances in controlled crystal growth and lithographic
techniques, it is now possible to fabricate various ``artificial 
atoms'' or quantum dots whose size is so small such that transport is 
in the ballistic regime\cite{review}. Among the many interesting 
phenomena associated with quantum ballistic transport, it was 
proposed\cite{marcus} that these quantum dots could be used to 
examine the theoretical notion of ``quantum 
chaos''\cite{gutzwiller,simons,ywang1}. More recently, Taylor 
{\it et.al.} fabricated an electronic Sinai billiard\cite{andy3} and 
investigated fluctuations of the resistance as an external magnetic
field is tuned. An important discovery of this work is the apparent
fractal resistance fluctuations\cite{fromhold} which maybe discussed on
the basis of a semiclassical theory\cite{ketzmerick}.

The problem of quantum chaos is a very interesting theoretical physics
issue, and our theoretical understanding of it has
been advanced by Random Matrix Theory (RMT) which classifies 
the statistics of the eigen-spectrum of a quantum 
system according to the Wigner-Dyson ensembles\cite{dyson}.  It is well 
established that, for a classically non-chaotic system such as a 
particle confined to move inside a rectangular box, the normalized 
nearest neighbor energy spacings $\{s\}$ obey a Poisson 
distribution\cite{berry2} $P(s)=e^{-s}$. On the other hand, 
for a classically chaotic system such as a spinless particle confined 
to move inside a stadium shaped box\cite{sinai,bunimovich}, 
the spacings follow the Wigner distribution 
which belongs to the Gaussian Orthogonal Ensemble (GOE) in the 
language of RMT\cite{bohigas1}. Furthermore, when time 
reversal invariance is broken, say by applying a magnetic field, 
the system is described\cite{mehta1} by Gaussian Unitary 
Ensemble (GUE), where 
$P_2(s)\approx\frac{32s^2}{\pi^2}e^{-\frac{4s^2}{\pi}}$.

In order to study certain aspects of quantum chaos using quantum 
{\it transport} techniques\cite{marcus,andy1,andy3}, one 
must deal with open systems where a scattering problem of charge 
carriers by some peculiar 
boundary must be solved. Experimentally it is quite difficult to study 
conduction as a function of the {\it incoming electron energy} in a 
quantitatively accurate fashion\cite{keller}, although measurements on
tunnel junctions made of Al nanoparticles have recently be made\cite{ralph}
and its relation to quantum chaos discussed\cite{agam}. Using 2D electron 
gas fabricated with compound semiconductors, experiments usually measure 
conductance as a function of external magnetic
fields\cite{marcus,andy1,andy3,berry}, $g(B)$, at a fixed 
electron Fermi energy $E_o$.
When a quantum dot is weakly coupled to the external leads, the 
magneto-conductance $g(B)$ may show resonance like behavior as the 
magnetic field $B$ is varied\cite{ywang1}, if the measurement is indeed
in a regime which probes the internal electronic states\cite{ralph}.
This behavior may be understood as follows. As $B$ is varied, the 
energy levels $\{\epsilon_i\}$ of the scattering states labeled by 
indices $i=1,2,...$ in the quantum dot change with it: 
$\epsilon_i=\epsilon_i(B)$. These levels are well separated since the
the dot region is weakly coupled to the leads. For an incoming 
electron with a fixed Fermi energy $E_o$, each time when the internal
state energy $\epsilon_i(B)$ is tuned to equal to $E_o$ as $B$ is 
varied, a resonance peak occurs in $g(B)$ due to a junction resonance. 
This was indeed observed in numerical simulations\cite{ywang1} by solving
the quantum scattering problem. Clearly this behavior should be observable
if the Coulomb blockade effects are small, which happens when the system
has a large capacitance thus small charging energy. We shall assume this
to be the case.

Hence in this junction resonance regime, it is interesting to define 
a ``B-spectrum'' as the collection of values of the special magnetic 
fields $B=\{B_i\}$ at which $g(B_i)$ is peaked\cite{ywang1}. It is 
important to ask the following questions: what are the statistical 
properties of this ``B-spectrum'' ?  How do its statistical properties 
change with the system shape ?  These are also useful questions to answer
because it is increasingly possible to directly probe the internal
electronic states of an isolated quantum dot, as demonstrated by the
experiment reported in Ref. \cite{ralph}.

Motivated by these questions, in this paper we report our studies on 
{\it closed} quantum dot systems where these questions can be answered 
clearly. Simons, Szafer and Altshuler\cite{simons} have investigated 
the correlations of slopes of the energy {\it levels} as a function 
of an external parameter such as the magnetic field $B$. We however 
emphasize that in their studies, the focus is on how a particular energy 
level $E_i$ varies with the external parameter $B$. Our investigation,
on the other hand, is completely different as it does not focus on 
any energy level: the energy $E_o$ is a fixed parameter in the Schr\"odinger 
equation by the equilibrium Fermi energy of the system. Rather,
we investigate the statistical properties of the set $\{B_i\}$ which makes
$E_o$ an eigenvalue of the Hamiltonian. Indeed, we note that fixing $B$ 
and study energy levels $\{E_i\}$, or fixing $E_o$ and study the 
``B-spectrum'' $\{B_i\}$, are two different problems. The former is about 
the energy eigenvalue spectra and its relation to an external parameter $B$,
while the latter is relevant for transport situations where the incoming
electron has energy $E_o$ which is fixed by the electron reservoir and 
cannot change, while one tunes $B$ to special values $\{B_i\}$ where 
conductance $g(B_i)$ is peaked. 
To our knowledge the statistical ensemble which is satisfied 
by the ``B-spectrum'' has not previously been determined. As we shall 
see below, this ``B-spectrum'' is determined by a {\it quadratic} 
eigenvalue problem, in distinct contrast to the usual energy spectrum 
at a fixed $B$ which is a linear eigenvalue problem. When a quantum 
dot is weakly coupled with the external leads, our answers are 
valid since in the weak coupling regime the levels are well separated 
and statistical properties should not change\cite{ywang,yan}.

To make the problem at hand clearer, the inset of Fig. (3) shows 
the two-dimensional quantum dot systems we have studied: a Sinai-like
billiard and a stadium-shaped quantum dot. These systems are classically 
chaotic systems. Experimentally transport measurements have been reported
through these quantum dots which were fabricated using the split-gate
technology\cite{andy1,andy3,marcus}. Fig. (1) shows the energy levels 
of a Sinai-like billiard as a function of magnetic field $B$.  For a 
given $B=B_o$, the levels satisfy GUE as mentioned above. These are the 
intersections of the line $B=B_o$ (the vertical solid line) with the 
spectrum.  However, for a fixed energy $E=E_o$, we are interested in the 
intersections of the horizontal solid line with the spectrum which define 
the ``B-spectrum''.  From the curves of Fig. (1) it is clearly not obvious 
what statistics the ``B-spectrum'' will satisfy.

In the presence of a magnetic field, the single-particle Schr\"{o}dinger
equation can be written as
\begin{equation}
\left\{ -\nabla^2-e_0 +b^2 A_0^2 +2b i \vec{A_0} \cdot \nabla \right\} 
\psi = 0\ \ \ ,
\label{eq4}
\end{equation}
where $\vec{A}=B\vec{A_0}=Bx\hat{j}$ is the vector potential,
$B$ is the magnetic field, $E = \hbar^2e_0/2m$ is the energy, 
$b=eB/\hbar c$, and $i^2 =-1$. Discretizing the spatial derivative, 
Eq.(\ref{eq4}) can be cast into a matrix form,
\begin{equation}
(M_1 +ibM_2 +b^2 M_3) \psi = e_0 \psi \ \ ,
\label{matrix_eq}
\end{equation}
where $M_1$, $M_2$, $M_3$ are real matrices and $M_3$ is also diagonal. 
From Eq.(\ref{matrix_eq}) it is clear that for a fixed magnetic field 
the solution of all the allowed energies forms the usual linear
eigenvalue problem which has been studied intensively.  
However for a fixed energy $e_0$, the solution of all the allowed magnetic 
fields $B$, the ``B-spectrum'', forms a quadratic eigenvalue problem.

There are two methods to find the ``B-spectrum''. The first method is by
brute force: one directly calculates the energy eigenvalues $E$ for a 
given magnetic field $B$ and traces out the curves of $E$ versus $B$. 
These curves may cross over each other. One then calculates, from 
this $E$ versus $B$ curve, the set of magnetic fields $B$ for a 
fixed energy $E_o$. Fig. (1) shows the $E$ versus $B$ curves for the 
first 40 eigenstates obtained using the Lanczos eigenvalue 
technique\cite{cullum} for the Sinai-like billiard system.  However, this 
method gets time consuming very quickly if higher and higher states 
are needed.

The second method is to transform the quadratic eigenvalue problem
Eq. (\ref{matrix_eq}) into a usual linear eigenvalue problem\cite{abo}. 
Let $\Psi(t) = e^{ibt} \psi$ and define $N_1 = M_3^{-1} 
(M_1 -e_0)$, $N_2 = M_3^{-1} M_2$, so that Eq. (\ref{matrix_eq}) becomes
$-I \Psi'' + N_2 \Psi' + N_1 \Psi = 0$
where $\Psi'=ib\Psi$ is the first derivative with respect to the 
parameter $t$, and $I$ is the unit matrix. With $\Psi' = \Phi$
($\Phi'=ib\Phi$), we have $\Phi = \Psi' = ib \Psi$ and 
$N_2 \Phi +N_1 \Psi = I \Phi'= ib \Phi$, or
\begin{equation}
\left( \begin{array}{cc} 0 & I \\ N_1 & N_2 \end{array} \right) 
\left( \begin{array}{c} \Psi \\ \Phi \end{array} \right) = ib 
\left( \begin{array}{c} \Psi \\ \Phi \end{array} \right)   .
\label{matrix_eq1}
\end{equation}
This is a linear eigenvalue problem for $b$. The matrix in 
Eq. (\ref{matrix_eq1}) is however not Hermitian, hence its eigenvalues 
may or may not be real. To obtain the physical solution, we 
look for all imaginary eigenvalues so that $b$ is real. 
We verified that the results coincide with those obtained with the
conventional Lanczos technique described above. In this way 
we have calculated the ``B-spectrum'' from Eq. (\ref{matrix_eq1}) 
for the two chaotic systems (inset of Fig. (3)), for different 
energies $e_0$.  The confining potential is assumed to be hard wall.

For the Sinai-like billiard, we considered a $4.7\mu m \times 4.3\mu m$
rectangular quantum dot with three hard disks inside the dot. The radius
of the disks were fixed at $0.9\mu m$, $0.8\mu m$, and $0.5\mu m$ 
with their centers randomly chosen but without the disks overlapping 
each other. Randomly changing positions of the disks allows us to
generate different configurations for ensemble averaging to obtain
reasonable statistics. The number of physical solutions
$N$ obtained from Eq.(\ref{matrix_eq1}) depends on the
fixed energy $e_0$, with $N$ increasing with the value of the energy
$e_0$. For instance, when $e_0$ is fixed at $30 meV$, we have $N\sim 1433$ 
physical solutions for the ``B-spectrum'' out of a total of about $3000$ 
eigenvalues. We found that the statistics of the ``B-spectrum''
behaves differently for the lower and higher part of the spectrum
respectively. The lowest\cite{foot1} 
$\sim 300$ ``B-levels'' give GUE statistics and the highest
$\sim 800$ ``B-levels'' give a Poisson-like behavior. Fig. (2) plots 
the distribution function obtained from our numerical data of the 
nearest neighbor ``B-level'' spacings for the Sinai-like billiards. With
the ensemble average of $20$ different configurations, the distribution 
determined from the lower part of the ``B-spectrum'' agrees well with  
GUE statistics (solid line) $P_2(s)$ discussed above.  On the other hand,
the higher part of the spectrum, shown in the inset of Fig. (2), has a 
Poisson-like behavior.

Another often used measure in studying level statistics is the 
spectral rigidity $\Delta_3$, defined as\cite{dyson,berry1} the mean 
square deviation of the best local fit straight line to the staircase 
cumulative spectral density over a normalized energy scale.  This 
quantity measures longer range correlations of the level spectrum and 
often provides a more critical test of the level statistics. To 
compute $\Delta_3$ we followed a scheme presented in Ref. \cite{casati}, 
and the numerical data is compared with the 
analytical formula from random matrix theory\cite{pandey}.  Fig (3) 
shows the $\Delta_3$ analysis of the lower part of the ``B-spectrum''. 
It is clear that the data is in very good agreement with the 
GUE statistics\cite{foot3}.

To test the statistical properties of the ``B-spectrum'' further, we 
studied another chaotic system, namely a stadium-shaped quantum 
dot (inset of Fig. (3)). The distribution function and $\Delta_3$ for
its ``B-spectrum'' are included in Figs. (2) and (3).  Here we 
did not use an ensemble average and the data is for one system only.
The general trend is the same as for the Sinai-like billiards, and still 
clearly shows the two distinct behaviors for different parts of 
the ``B-spectrum'',  namely a GUE behavior of the lower part 
and a Poisson-like behavior for the higher part. Finally, 
we have verified that the same statistical behavior is observed for 
both the Sinai-like billiard and the stadium-shape billiard with many 
different values of the fixed energy $e_0$, and conclude that the 
lower part of the ``B-spectrum'' of both systems satisfies 
GUE statistics.

It is not difficult to understand that the higher part of the
``B-spectrum'' should behave differently.  The higher part corresponds
to larger values of the magnetic fields, which are known to destroy 
chaos\cite{robnik,gutzwiller}.  For the particular sizes of the 
Sinai-like billiard and
the stadium-shaped dot, the magnetic-field $B_c$ that roughly separates 
the low lying and high lying part of the ``B-spectrum'' is about 
$B_c \sim 3$ to $4$T.  The classical cyclotron radius of the electron 
at and above this field strength is quite small compared with the system 
size. Hence the electron ``skips'' along the wall of the confining
potential or makes circular motion inside the quantum dot,
thereby reducing the effect of chaotic scattering by the
geometry\cite{foot2}. Nevertheless, the field range up to $B_c$ is quite
wide and it should be possible to investigate the ``B-spectrum'' 
experimentally using resonant magneto-conductance measurements for 
systems having weak coupling to the leads.

While the quadratic eigenvalue problems reported above were for
billiard systems where the continuum Schr\"odinger equation is
discretized to obtain the matrix equation (\ref{matrix_eq1}),
we also studied a similar quadratic eigenvalue problem using
random matrices to replace the matrices $M_1$, $M_2$ and $M_3$ in Eq.
(\ref{matrix_eq}).  Two requirements must be satisfied:  first,
the matrices $M_1$, $M_2$, and $M_3$ must be real; second, $M_1$ must
be symmetric and $M_2$ antisymmetric so that the Hamiltonian be 
Hermitian. The random matrices were set up in the standard fashion: for
a $N \times N$ matrix, $M_1$ has $N(N+1)/2$ independent matrix elements
and $M_2$ has $N(N-1)/2$, where the matrix elements are Gaussian 
random numbers.  We then diagonalized Eq.(\ref{matrix_eq1}) using 
the same numerical methods discussed above and found all the physical 
solutions.  We chose $N=1500$ so that the matrix which must be 
diagonalized is $3000 \times 3000$. Typically we obtained about 
700 physical ``B-levels'' for various energies $e_0$. The results 
are included in Figs. (2) and (3). It is clear that the ``B-level'' 
statistics from the random matrices is completely consistent with the 
GUE statistics. We may conclude that ``B-spectra'' coming from the 
quantum billiard systems and from random matrices are still consistent 
with each other and a universality can still be established for this 
quadratic eigenvalue problem using the random matrices.

In summary, we have numerically investigated the statistical properties
of the magnetic field spectra (the ``B-spectrum''), which is determined 
by a quadratic eigenvalue problem. This spectrum is defined by the 
allowed magnetic fields for an electron moving in a quantum dot with 
its fixed fermi energy.  This problem arises for systems with weak 
coupling to the leads in which the scattering states are well separated 
in energy. For two different chaotic billiards, {\it e.g.} two-dimensional 
quantum dots in the shape of a Sinai-like billiard
and a stadium billiard, the ``B-spectra'' have distinctly different 
statistical behavior at the lower and higher parts of the spectra. 
In particular, the lower part is well described by the GUE statistics
while the higher part is Poisson like.  We found that the same quadratic
eigenvalue problem can be studied using random matrices as well, 
and the eigenvalues from the random matrices have precisely the
same behavior as those of the billiards. Thus the notion of universality 
classes using the random matrix theory can be carried over to this 
new problem. While our numerical data provided clear evidence of the 
statistical properties of the present quadratic eigenvalue problem, 
it is not at all obvious {\it a priori} that such statistical properties 
are controlled by the GUE universality.  
Further work is needed to provide an analytical understanding.

Experimentally the statistical properties of the ``B-spectra'' could be
examined for quantum dot systems which couple weakly to the external
leads and thus the transmission is controlled by junction resonances.
The weak coupling could be provided by adding constrictions at the
connections of the leads with the quantum dot for which only the
lowest few subbands of the leads can propagate. In typical experimental
situation on submicron structures the single particle level spacing is
around 0.05meV, thus they can be measured if the temperature is 
kept to less than 500mK. We thus conclude that for a magneto-transport 
measurement in the junction resonance regime, the special magnetic
field strengths $\{B_i\}$ at which $g(B_i)$ takes extremal values, should
satisfy GUE statistics.

\section*{Acknowledgments}

We thank Prof. R. Harris for useful discussions and a critical 
reading of the manuscript. We gratefully acknowledge support by 
a research grant from the Croucher Foundation, a RGC grant from the 
Government of Hong Kong under grant number HKU 261/95P, 
the Natural Sciences and Engineering Research Council 
of Canada and le Fonds pour la Formation de
Chercheurs et l'Aide \`a la Recherche de la Province du Qu\'ebec.
We also thank the Computer Center of the University of Hong Kong for
computational facilities.

\section*{Figure Captions}

\begin{itemize}

\item[{Figure 1.}] Typical curves of the energy levels as a function of
the magnetic field, $E=E(B)$, for the Sinai-like
billiard.  The vertical 
solid line at $B=B_o$ intersects the spectrum and the intersections
give the usual energy levels $E=E(B_o)$.  The ``B-spectrum'' is the 
collection of intersections of the horizontal solid line at a 
fixed energy $E=E_o$.

\item[{Figure 2.}] The distribution function of the nearest neighbor
``B-level'' spacings obtained from the lower part of the 
``B-spectrum". The solid line is the analytical formula for 
GUE $P_2(s)$. Solid squares: ensemble averaged data for $20$ 
independent Sinai-like billiards; open circles: for a stadium-shaped 
quantum dot; triangles: for random matrices averaged over 
13 configurations.  Insert: the distribution function obtained from
the higher part of the ``B-spectrum'' of the same systems
(symbols) and the solid line is the Poisson distribution $P(s)$. 
The energy is fixed at $e_0=30meV$.

\item[{Figure 3.}] The spectral rigidity $\Delta_3(L)$, where $L$ is the 
number of the ``B-levels'' involved in computing $\Delta_3$.
The solid line is the analytical expression for GUE statistics, 
taken from Ref.[18].  Solid squares: ensemble averaged data for 
$20$ independent Sinai-like billiards; open circles: for a stadium-shaped 
quantum dot; triangles: from the random matrices averaged over 
13 configurations.  Insert: sketch of the chaotic 
structures studied here, a Sinai-like billiard with three hard disks confined 
inside a rectangle; and a stadium-shaped quantum dot. 

\end{itemize}


\newpage

\begin{thebibliography}{00}

\bibitem{review}
For a review of mesoscopic physics, see C.W.J. Beenakker and H. van 
Houten, Solid State Phys. {\bf 44}, 1(1991) and references therein;
M.A. Kastner, Physics Today, {\bf 46}, 24 (1993).

\bibitem{marcus}
C. M. Marcus, A. J. Rimberg, R. M. Westervelt, P. F. Hopkins, and A. C.
Gossard, Phys. Rev. Lett. {\bf 69}, 506 (1992); C.M. Marcus, R.M.
Westervelt, P.F. Hopkins and A.C. Gossard, Phys. Rev. B {\bf 48}, 2460
(1993).

\bibitem{gutzwiller}
M. Gutzwiller, Chaos in Classical and Quantum Mechanics (Springer Verlag, New
York, 1991). 

\bibitem{simons}
B.D. Simons, A. Szafer, and B.L. Altshuler, JETP Lett. {\bf 57}, 276 (1993);
A. Szafer and B.L. Altshuler, Phys. Rev. Lett. {\bf 70}, 587 (1993);
B. D. Simons and B. L. Altshuler, Phys. Rev. Lett. {\bf 70}, 4063 (1993).

\bibitem{ywang1}
Y.J. Wang, J. Wang, H. Guo, and C. Roland, J. Phys. Condens. 
Matter {\bf 6}, L143 (1994).

\bibitem{andy3}
R.P. Taylor, R. Newbury, A.S. Sachrajda, Y. Feng, 
P.T. Coleridge, C. Dettmann, Ningjia Zhu, Hong Guo, A. Delage, 
P.J. Kelly and Z. Wasilewski, Phys. Rev. Lett., {\bf 78}, 1952 (1997).

\bibitem{fromhold}
For a short review of the fractal resistance fluctuations observed in Ref.
\cite{andy3}, see, Mark Fromhold, Nature, {\bf 386}, 123 (1997).

\bibitem{ketzmerick}
Roland Ketzmerick, Phys. Rev. B. {\bf 54}, 10841 (1996); A.S. Sachrajda
{\it et. al.}, Cond-mat/9709091.

\bibitem{dyson}
F. J. Dyson, J. Math. Phys. {\bf 3}, 140 (1962); ibid, 157 (1962); 
ibid 166 (1962); F. J. Dyson and M. L. Mehta, J. Math. Phys. {\bf 4}, 
701 (1963); M. L. Mehta and F. J. Dyson, ibid, {\bf 4}, 713 (1963). 

\bibitem{berry2}
M.V. Berry and M. Tabor, Proc. Roy. Soc. London, A {\bf 356}, 375(1977).

\bibitem{sinai}
Y.G. Sinai, Fanct. Anal. Appl. {\bf 2}, 61 (1968); Russ. Math. Surveys
{\bf 25}, 137 (1970).

\bibitem{bunimovich}
L.A. Bunimovich, Funct. Anal. Appl. {\bf 8}, 254 (1974); Commun. Math.
Phys. {\bf 65}, 295 (1979); L.A. Bunimovich and Y.G. Sinai, Commun.
Math. Phys. {\bf 78}, 247 (1980).

\bibitem{bohigas1}
For a general review, see O. Bohigas, in Proceedings of the Les
Houches Summer School, Session LII (Elsevier Science Publishers, New
York, 1989). 

\bibitem{mehta1}
A. Pandey and M.L. Mehta, Commun. Math. Phys., {\bf 87}, 449(1983);
M.L. Mehta and A. Pandey, J. Phys. A, {\bf 16}, 2655 (1983);
J.B. French, V.K.B. Kota, A. Pandey, and S. Tomsovic, Ann. Phys., {\bf 181},
198 (1988).

\bibitem{andy1}
R.P. Taylor, R. Newbury, A.S. Sachrajda, Y.  Feng, P.T. Coleridge, 
A. Delage, P.J. Kelly, Z. Wasilewski, P. Zawadzki, Ningjia Zhu and 
Hong Guo, Superlattics and Microstructures, {\bf 20}, 297 (1996).

\bibitem{keller}
M. W. Keller, O. Millo, A. Mittal, D. E. Prober, and R. N. Sacks, Surf.
Sci. {\bf 305}, 501 (1994). 

\bibitem{ralph}
D.C. Ralph, C.t. Black and M. Tinkham, Phys. Rev. Lett. {\bf 78}, 4087 
(1997).

\bibitem{agam}
Oded Agam, Ned S. Wingree, Boris L. Altshuler, D.C. Ralph and M. Tinkham,
Phys. Rev. Lett. {\bf 78}, 1956 (1997).

\bibitem{berry}
M. J. Berry, J. H. Basky, R. M. Westervelt, and A. C. Gossard, Phys.
Rev. B {\bf 50}, 8857 (1994); M. J. Berry, J. A. Katine, C. M. Marcus,
R. M. Westervelt, and A. C. Gossard, Surf. Sci. {\bf 305}, 480 (1994);
I. H. Chan, R. M. Clarke, C. M. Marcus, K. Campman, and A. C. Gossard,
Phys. Rev. Lett. {\bf 74}, 3876 (1995). 

\bibitem{ywang}
J. Wang, Y. J. Wang, and H. Guo,  Appl. Phys. Lett. {\bf 65}, 1793 (1994); 
Y. J. Wang, N. J. Zhu, J. Wang, and H. Guo, Phys. Rev. B {\bf 53}, 16408
(1996). 

\bibitem{yan}
``Parametric Correlations of Phase Shifts and Statistics of Time Delays in 
Quantum Chaotic Scattering: Crossover between Unitary and Orthogonal 
Symmetries'', Yan V. Fyodorov,  Dmitry V. Savin, and Hans-J\"{u}rgen Sommers,
preprint, (1997).

\bibitem{cullum}
J. Cullum and R. A. Willoughby, J. Compt. Phys. {\bf 44}, 329 (1981). 

\bibitem{abo}
M. Abo-Hamd et al, Journal of the Engineering Mechanics Division 
{\bf 104}, 537 (1978). 

\bibitem{foot1}
For $B\approx 0$ the energy levels $E(B)$ satisfy GOE statistics, and it
changes to GUE when $B$ is such that approximately one flux quanta is
enclosed by the quantum dot area.  In our case here, below this
transition field there are about 16 to 17 ``B-levels''.  Thus in all our 
data analysis, we have not included the lowest twenty ``B-levels''. 

\bibitem{berry1}
M.V. Berry, Proc. Roy. Soc. (London), A {\bf 400}, 229 (1985).

\bibitem{casati}
G. Casati, B.V. Chirikow and I. Guarneri, Phys. Rev. Lett. {\bf 54}, 
1350 (1985).

\bibitem{pandey}
A. Pandey, In {\it Quantum Chaos and Statistical Nuclear Physics}, 
Lecture notes in Physics, {\bf 263}, E. T.H. Seligman and H. Nishioka, 
(Springer-Verlag, Berlin, 1986), p98.

\bibitem{foot3}
A departure from random matrix prediction was observed in Fig. (3) when
$L$ becomes larger than $20$.  This is similar to that happens in energy
level analysis.

\bibitem{robnik}
M. Robnik, J. Phys. A {\bf 14}, 3195 (1981).

\bibitem{foot2}
For the high lying part of the ``B-spectrum'', the precise distribution
form is difficult to determine from our numerical analysis. However a
Poisson-like behavior, namely a ``B-level'' attraction, is clearly
observed here.  The numerical fit to a Poisson distribution is 
nevertheless not good, due in part to two reasons. First, there is 
a gradual crossover from the GUE distribution of the low lying part of 
the spectrum.  Second, for very high part of the ``B-spectrum'', the
electron wavefunction forms edge states with nearly equal level
spacings (the Landau levels) where the energy versus $B$ is nearly
linear. These lead to a deviation from a perfect Poisson distribution.

\end{thebibliography}
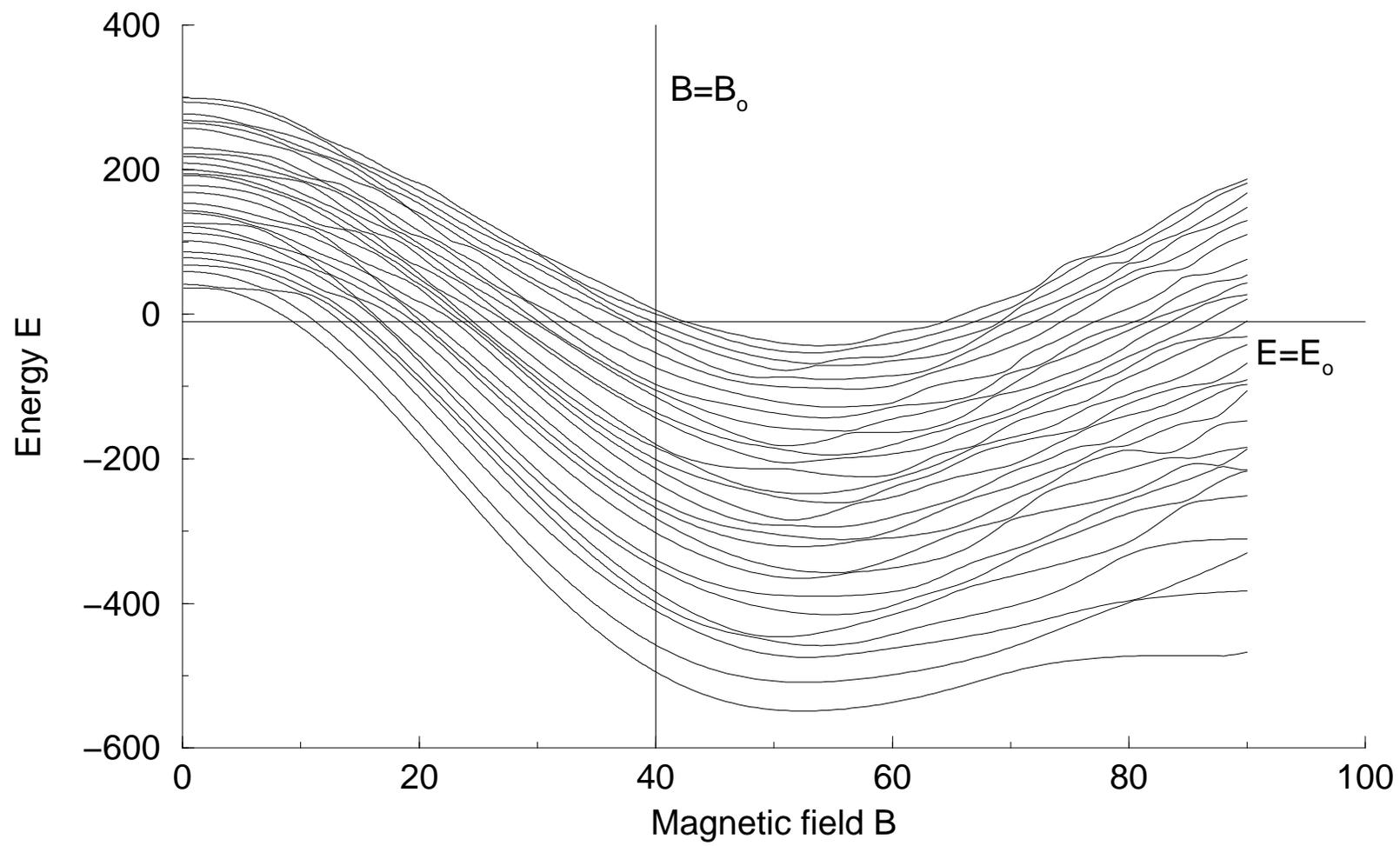
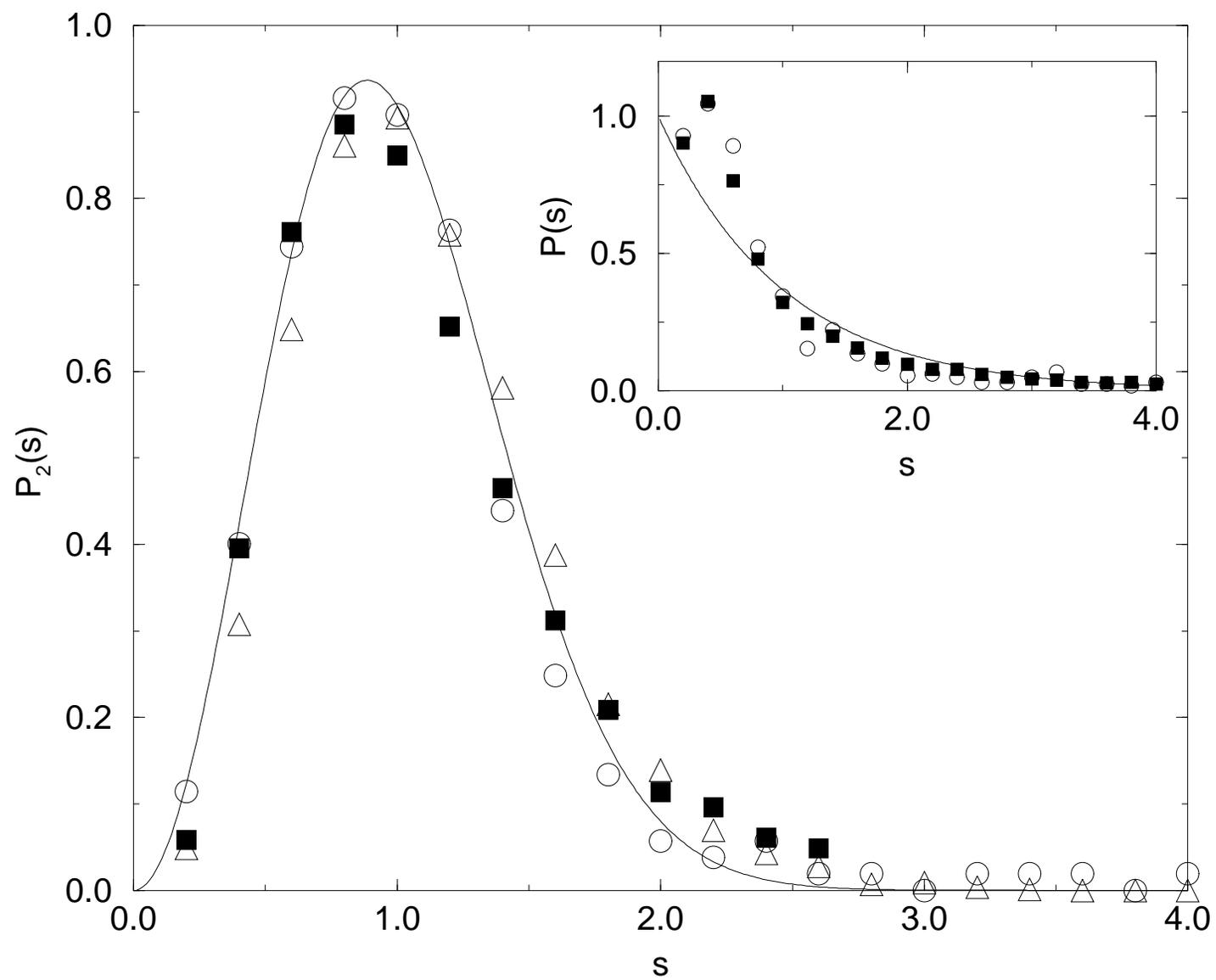
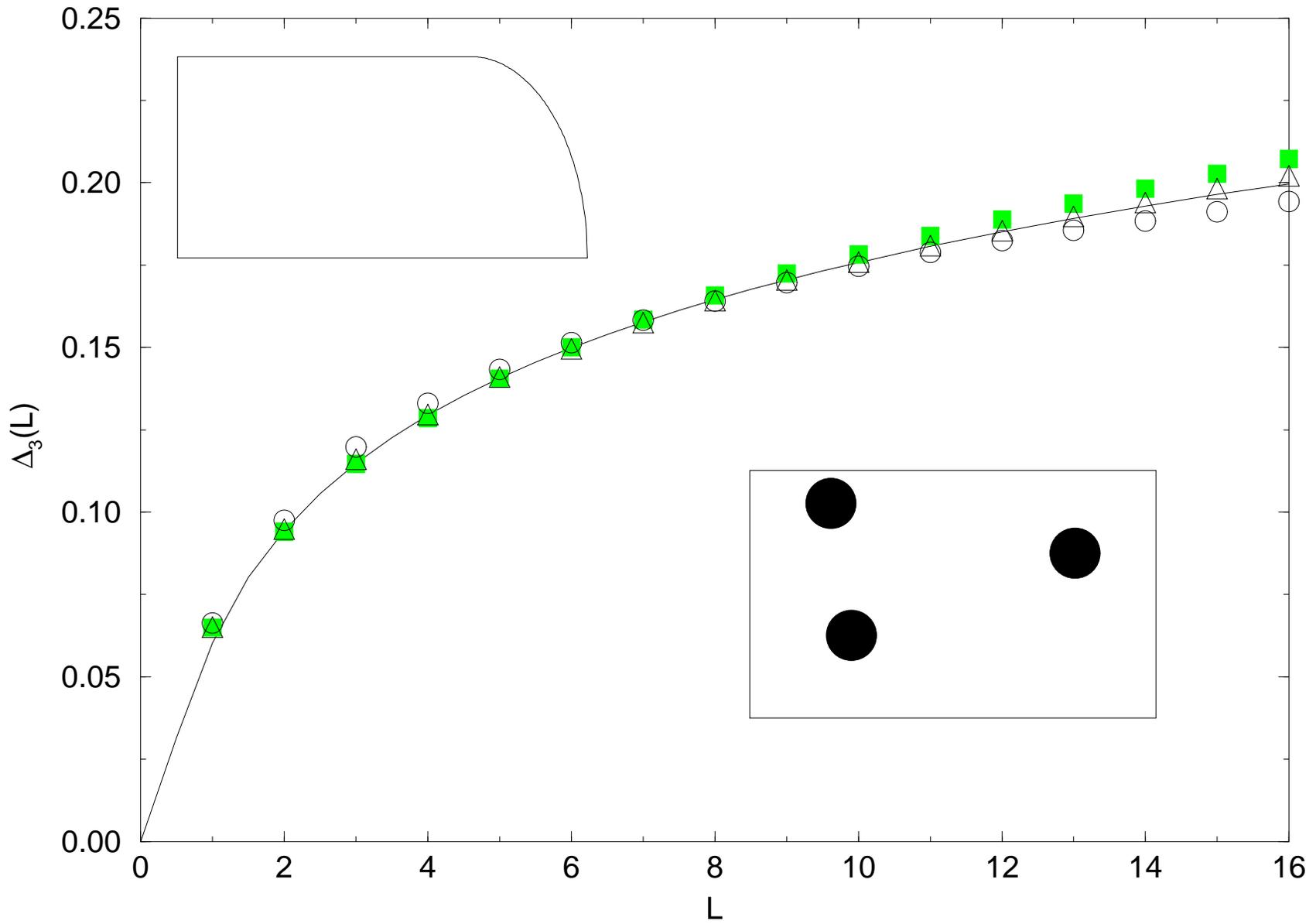
\end{document}